\documentclass[pre,showpacs,preprint,superscriptaddress]{revtex4}
\usepackage{amssymb,amscd,amsmath}
\usepackage{graphicx}
\usepackage{color}
\usepackage{array,calc,multirow,mathtools}
\def\hsp{\hspace}

\def\tm{truemm}
\def\a={&=&}
\def\aa={&\approx&}
\def\ad={&\equiv&}
\def\ap={&\propto&}
\def\as= {&\sim&}
\def\lra={&\Leftrightarrow&}
\def\sa={\hsg &=& \hsg}
\def\saa={\hsg &\approx&\hsg}
\def\sad={\hsg &\equiv&\hsg}
\def\sap={\hsg &\propto&\hsg}
\def\sass={\hsg &\sim&\hsg}
\def\slra={\hsg &\Leftrightarrow&\hsg}
\def\b={\ = \ }
\def\ba={\ \approx \ }
\def\bp={\ \propto \ }
\def\c={\hsa = \hsa}
\def\ca={\hsa \approx \hsa}
\def\abs= {\ \sim \ }
\def\acs= {\hsa \sim \hsa}

\def\bHu{\begin{Huge}}
\def\bhu{\begin{huge}}
\def\bLA{\begin{LARGE}}
\def\bLa{\begin{Large}}
\def\bla{\begin{large}}
\def\bsm{\begin{small}}
\def\bft{\begin{footnotesize}}
\def\bsc{\begin{scriptsize}}
\def\bti{\begin{tiny}}
\def\eHu{\end{Huge}}
\def\ehu{\end{huge}}
\def\eLA{\end{LARGE}}
\def\eLa{\end{Large}}
\def\ela{\end{large}}
\def\esm{\end{small}}
\def\eft{\end{footnotesize}}
\def\esc{\end{scriptsize}}
\def\eti{\end{tiny}}
\def\barr{\begin{array}}
\def\earr{\end{array}}
\def\bc{\begin{center}}
\def\ec{\end{center}}
\def\ben{\begin{enumerate}}
\def\een{\end{enumerate}}
\def\beq{\begin{equation}}
\def\eeq{\end{equation}}
\def\beqa{\begin{eqnarray}}
\def\eeqa{\end{eqnarray}}
\def\beqanu{\vspace{-7mm} \begin{eqnarray*}}
\def\beqan{\begin{eqnarray*}}
\def\eeqan{\end{eqnarray*}}
\def\bit{\begin{itemize}}
\def\eit{\end{itemize}}
\def\bpm{\begin{pmatrix}}
\def\epm{\end{pmatrix}}
\def\bvs{\begin{verse}}
\def\evs{\end{verse}}
\def\btab{\begin{tabular}}
\def\etab{\end{tabular}}

\def\hsa{\mbox{} \hspace{2\tm} }
\def\hsb{\mbox{} \hspace{4\tm} }

\def\hsg{\mbox{} \hspace{-2\tm} }

\newcommand{\av}[1]{\overline{#1}}
\newcommand{\ave}[1]{\langle {#1} \rangle}

\newcommand{\mf}{\hspace{5mm} \mbox{for} \ \ }

\begin{document}
\title{Duality between cooperation and defection in the presence of tit-for-tat
in replicator dynamics}

\author{Seung Ki Baek}
\email[]{seungki@pknu.ac.kr}
\affiliation{Department of Physics, Pukyong National University, Busan 48513,
Korea}

\author{Su Do Yi}
\email[]{esudoz@snu.ac.kr}
\affiliation{CCSS, Department of Physics and Astronomy, Seoul National
University, Seoul 08826, Korea}

\author{Hyeong-Chai Jeong}
\email[]{hcj@sejong.edu}
\affiliation{Department of Physics and Astronomy, Sejong University, Seoul
05006, Korea}
\affiliation{Quantum Universe Center, Korea Institute for Advanced Study, Seoul
02455, Korea}

\begin{abstract}
The prisoner's dilemma describes a conflict between a pair of players,
in which defection is a dominant strategy whereas cooperation is collectively
optimal. The iterated version of the dilemma has
been extensively studied to understand the emergence of cooperation.
In the evolutionary context, the iterated prisoner's dilemma is often
combined with population dynamics, in which a more successful strategy
replicates itself with a higher growth rate. Here, we investigate the replicator
dynamics of three representative strategies, i.e., unconditional cooperation,
unconditional defection, and tit-for-tat, which prescribes reciprocal
cooperation by mimicking the opponent’s previous move. Our finding is that the
dynamics is self-dual in the sense that it remains invariant when we apply time
reversal and exchange the fractions of unconditional cooperators and defectors
in the population. The duality implies that the fractions can be equalized by
tit-for-tat players, although unconditional cooperation is still dominated by
defection. Furthermore, we find that mutation among the strategies breaks the
exact duality in such a way that cooperation is more favored than defection, as
long as the cost-to-benefit ratio of cooperation is small.
\end{abstract}

%02.50.Le 	Decision theory and game theory
%87.23.Cc 	Population dynamics and ecological pattern formation
%05.45.-a	Dynamical systems, nonlinear


\maketitle

\section{Introduction}

Although a society consists of individuals, the collective interest is not an
aggregate of individual ones. The prisoner's dilemma (PD) game is a toy model to
illustrate such a social dilemma. The PD game can be formulated as follows:
Suppose that we have two players, say, Alice
and Bob. When Alice cooperates, it benefits Bob by a certain amount of $b$ at
her own cost $c$. If she defects, on the other hand, it does not incur any cost
and Bob gains nothing.
If $c$ exceeds $b$, defection obviously drives out cooperation, so we restrict
ourselves to $0<c<b$. The cost-to-benefit ratio, $c/b$, is thus limited to an
open interval $(0,1)$.
The resulting payoff matrix between cooperation (C) and
defection (D) is expressed as
\beqa
\hsp{-10\tm}
\begin{array}{cl}
 \hsb   & \begin{array}{cc}
     \hsa\hsb\hsa  C  &\hsb\hsb D
       \end{array} \ \	\\
\begin{array}{c}
      C      \\
      D
\end{array}
   & \left( \begin{array}{ccc}
\hsa    b-c   & \hsb  -c\\
\hsa    b     & \hsb   0
  \end{array} \right),
\end{array}
\label{e.CD}
\eeqa
from the row-player Alice's point of view, and the game is symmetric to both
players. The collective interest is maximized when both choose $C$, but $D$ is
the rational choice for each individual, hence a dilemma.

By construction of the PD game, unconditional defection (AllD) always
constitutes a Nash equilibrium.
However, it has been widely known by folk theorems that
a cooperative strategy can also be rational if the PD game is repeated
{\it indefinitely} with high enough probability
because one's cooperation can be reciprocated by the other's
in future. This is called direct reciprocity and has been popularized
by Axelrod's tournament of the iterated prisoner's dilemma
(IPD)~\cite{axelrod:book:1994}. We assume that the repetition probability
approaches one. An archetypal strategy of direct reciprocity
is Tit-for-tat (TFT). It begins with $C$ at the first encounter and then
replicates the co-player's last move. Except the first round, therefore, it
cooperates only if the co-player cooperated last time. We may call it a
conditional cooperator, opposed to an unconditional cooperator (AllC).
We will explain that the interactions between the aforementioned strategies, i.e., AllD, TFT, and AllC, are rather subtle, indicating the complexity in
evolution of cooperation. Earlier studies have already focused on the dynamics
of these three representative
strategies~\cite{imhof:pnas:2005,brandt:jtb:2006,toupo:ijbc:2014}.

All these fall into a class of reactive strategies~\cite{baek:scirep:2016}
represented by a two-component array
$\alpha = (P_C, P_D)$, where $P_C$ ($P_D$) means the probability to cooperate
when the co-player cooperated (defected) last time. In this notation, we have
AllC = $(1,1)$, AllD = $(0,0)$, and TFT = $(1,0)$. If error occurs with
probability $e$ at each time step, the effective behavior is described as
$\alpha' = ((1-e)P_C + e(1-P_C), (1-e)P_D + e(1-P_D)) = (P'_C, P'_D)$.
The error rate $e$ is assumed to be small, and this statement will be made
quantitative later.
Suppose that two strategies $\alpha = (P_C,
P_D)$ and $\beta = (Q_C, Q_D)$ meet in the IPD. They effectively behave as
$\alpha'$ and $\beta'$, respectively, and stochastically visit four
states, $CC, CD, DC$, and $DD$, where the former (latter) symbol means the move
of the player adopting $\alpha$ ($\beta$). The transition probabilities between
the states can be arranged in the following
matrix~\cite{nowak:jtb:1989,nowak:aam:1990}:
\begin{equation}
\tilde{M} = \left(
\begin{array}{cccc}
P'_C Q'_C & P'_D Q'_C & P'_C Q'_D & P'_D Q'_D\\
P'_C (1-Q'_C) & P'_D (1-Q'_C) & P'_C (1-Q'_D) & P'_D (1-Q'_D)\\
(1-P'_C) Q'_C & (1-P'_D) Q'_C & (1-P'_C) Q'_D & (1-P'_D) Q'_D\\
(1-P'_C) (1-Q'_C) & (1-P'_D) (1-Q'_C) & (1-P'_C) (1-Q'_D) & (1-P'_D) (1-Q'_D)
\end{array}
\right).
\end{equation}
This stochastic matrix is irreducible and positive definite, so the
Perron-Frobenius theorem guarantees the existence of a unique right eigenvector
$\vec{v} = (v_{CC}, v_{CD}, v_{DC}, v_{DD})$ with the largest eigenvalue
$\Lambda = 1$. If we normalize $\vec{v}$ in such a way that $v_{CC} + v_{CD} +
v_{DC} + v_{DD} = 1$, it is the stationary probability distribution over the
four states when the strategies $\alpha$ and $\beta$ are adopted in the IPD. The
long-term payoff of $\alpha$ against $\beta$ per round is obtained by
calculating an inner product $p_{\alpha \beta} = \vec{v} \cdot \vec{h}_1$, where
$\vec{h}_1 = (b-c, -c, b, 0)$.  Likewise, we obtain $p_{\beta \alpha} = \vec{v}
\cdot \vec{h}_2$ with $\vec{h}_2 = (b-c, b, -c, 0)$. If we list the three
strategies in the order of AllC, AllD, and TFT, the matrix $\tilde{p} =
\{ p_{\alpha \beta}\}$ can be written as follows:
\begin{equation}
\tilde{p} =
\begin{pmatrix}
(b-c)(1-e) & be-c(1-e) & b(1-2e+2e^2) - c(1-e)\\
b(1-e)-ce & (b-c)e & 2b(1-e)e-ce \\
b(1-e)-c(1-2e+2e^2) & be-2c(1-e)e & (b-c)/2
\end{pmatrix}.
\label{e.pab}
\end{equation}
Note that the limit of $e \rightarrow 0$ does not coincide with the case of
$e=0$: If $e$ was strictly zero between two TFT players, each of them would earn
$b-c$ at each round. For any $e > 0$, however, the average payoff per round
reduces to $(b-c)/2$ as written in Eq.~\eqref{e.pab}.
All these results are fully consistent with existing ones such as in
Refs.~\onlinecite{molander:jcr:1985,imhof:jtb:2007}.

In an evolutionary framework, we consider dynamics of a well-mixed population
in which random pairs of individuals play the IPD game. Let us assume that the
population is so large that stochastic fluctuations can be
ignored. If a certain strategy earns a higher payoff than the population
average, we can expect that its fraction will grow at a rate proportional to
the payoff difference from the population average. Likewise, a strategy with a
lower payoff than the population average will decrease in its fraction.
Replicator dynamics (RD) expresses this idea by using a set of deterministic
equations for the time evolution of the fractions.
Let $N_s$ be the total number of strategies in the population. We have $N_s=3$
in a set of the three strategies, i.e., \{AllC, AllD, TFT\}.
We are interested in the fraction $x_\alpha$ of strategy $\alpha$, with a
normalization condition that $\sum_\alpha x_\alpha = 1$.
The long-term payoff of strategy $\alpha$ from the whole
population is denoted as
\begin{equation}
p_\alpha = \sum_\beta p_{\alpha \beta} x_\beta.
\label{e.pa}
\end{equation}
RD describes the time evolution of $x_\alpha$ as follows:
\beqa
\frac{dx_\alpha}{dt} \a= \sum_\beta q_{\alpha \beta} p_\beta x_\beta - \ave{p}
x_\alpha,
\label{e.rd0}
\eeqa
where $q_{\alpha \beta}$'s are elements of a transition matrix between
strategies. The average payoff of the population is denoted as $\ave{p}
\equiv \sum_\alpha p_\alpha x_\alpha = \sum_{\alpha \beta} p_{\alpha \beta}
x_\alpha x_\beta$. If we choose the transition matrix as
\beqa
 q_{\alpha\beta} \a= \left\{\barr{ll}
             1-\mu  & \mf \ \alpha = \beta \\
   \mu/(N_s-1)& \mf \ \alpha\not= \beta,
   \earr
   \right.
\label{e.qab}
\eeqa
RD takes the following form:
\beqa
\frac{d x_\alpha}{dt} \a= (1-\mu) p_\alpha x_\alpha - \ave{p} x_\alpha +
\frac{\mu}{N_s-1} \sum_{\beta\not=\alpha} p_\beta x_\beta,
\label{e.rd}
\eeqa
where $\mu$ is a mutation rate, assumed to satisfy $\mu \ll e$.
The first term on the right-hand side means growth with a rate proportional to
the payoff, the second term normalizes the total sum of $x_\alpha$'s, and the
last term describes mutation. Note that the fitness of strategy $\alpha$ is
identified with its payoff $p_{\alpha} (t)$, so that it produces offspring in
proportion to $p_{\alpha} (t) x_{\alpha} (t)$ between time $t$ and $t+dt$.
The mutation structure in Eq.~\eqref{e.qab} means that some of these offspring
are randomly picked up and change the strategy to one of the others.

In this work, we will show the following:
If $\mu$ vanishes, the time evolution of $x_{\mbox{\tiny AllC}}$ in RD is
the same as that of $x_{\mbox{\tiny AllD}}$ under time reversal, $t \rightarrow
-t$, and vice versa. The duality does not exactly hold for $\mu > 0$, and
we will discuss its consequences by analyzing the system perturbatively.

\section{Fixed-point Structure}

\begin{figure}
\includegraphics[width=0.45\textwidth]{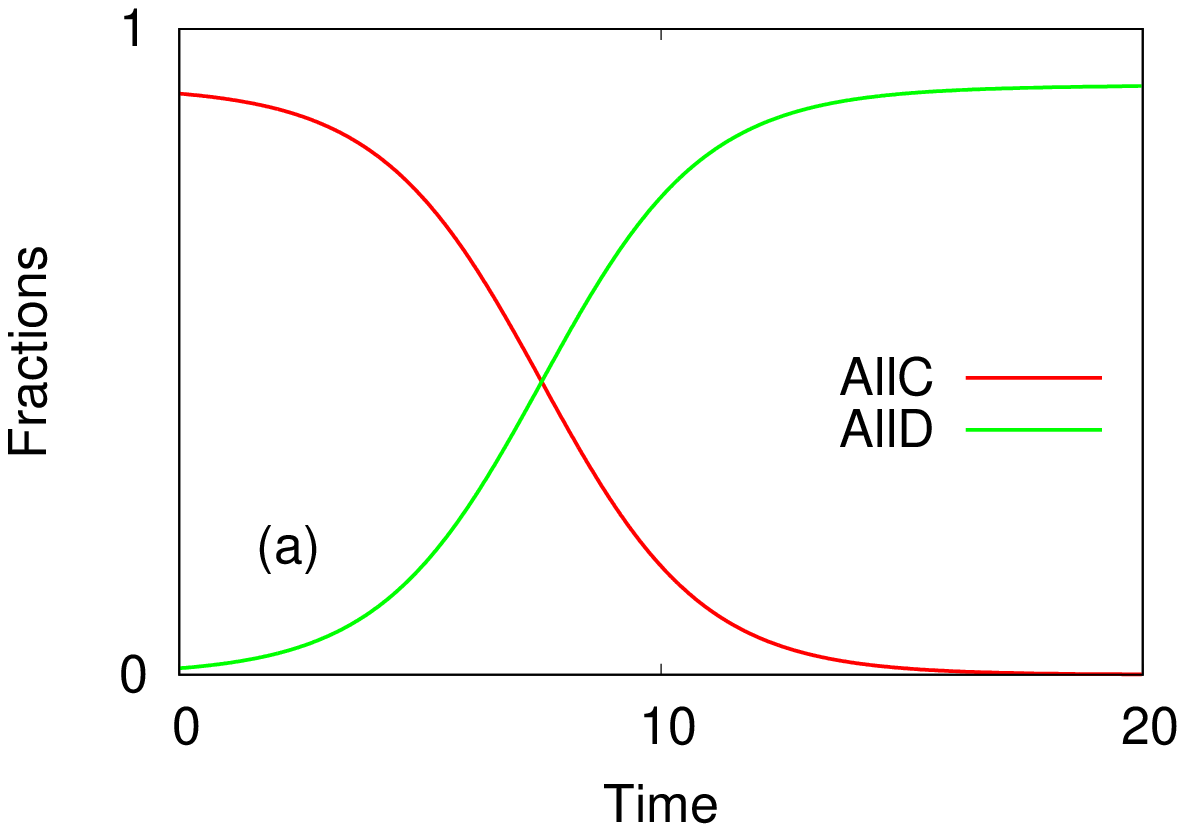}
\includegraphics[width=0.45\textwidth]{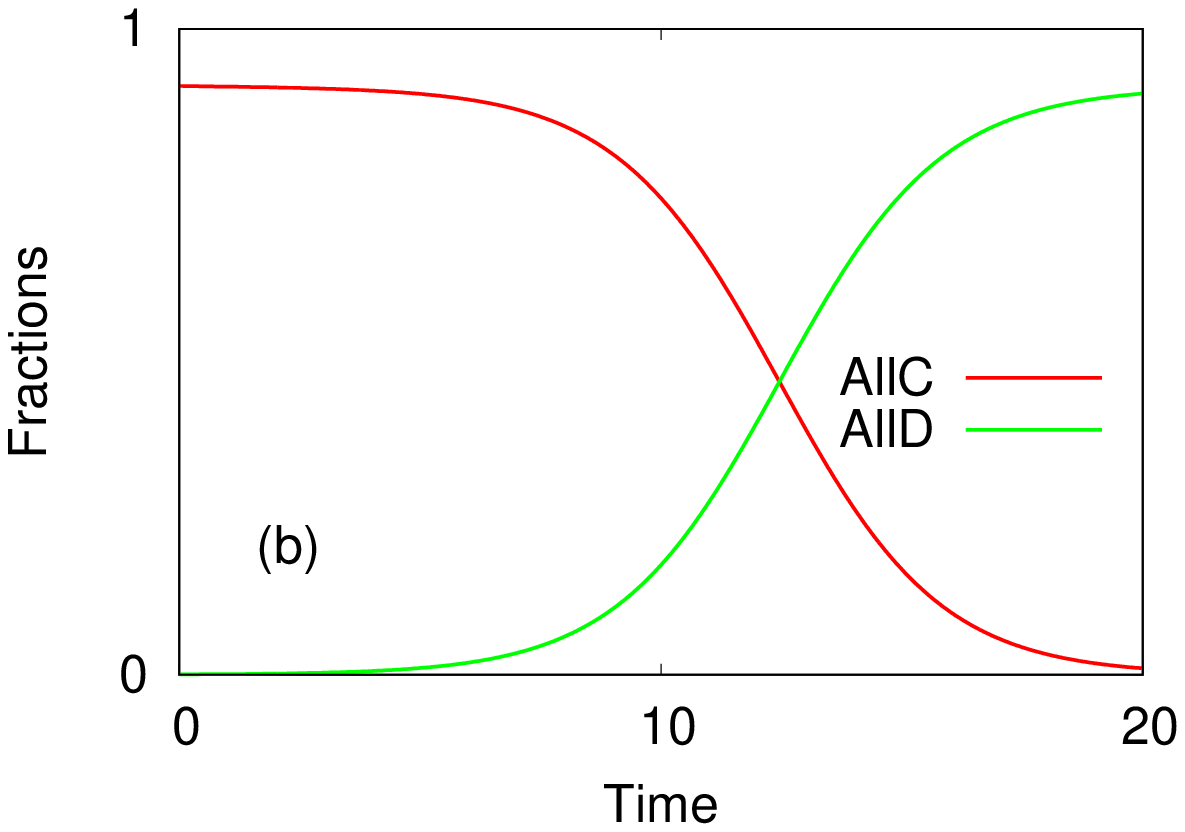}
\caption{Explicit example of duality. (a) An evolutionary trajectory resulting
from the mutation-free RD. (b) A mirror image of the left panel upon time
reversal and exchange between AllC and AllD. At the same time, it shows a
completely legitimate trajectory under the same dynamics.}
\label{f.dual}
\end{figure}

For the sake of notational convenience, we define $x_1 \equiv x_{\mbox{\tiny
AllC}}$, $x_2 \equiv x_{\mbox{\tiny AllD}}$, and $x_3 \equiv x_{\mbox{\tiny
TFT}}$ henceforth. Due to the normalization condition, we have only two
independent variables, which we choose as $x_1$ and $x_2$. Plugging
Eq.~\eqref{e.pa} into Eq.~\eqref{e.rd}, we find a set of equations, which can
be formally written as follows:
\begin{eqnarray}
\frac{dx_1}{dt} &=& f_1(x_1, x_2; e, \mu)\label{e.rd1}\\
\frac{dx_2}{dt} &=& f_2(x_1, x_2; e, \mu)\label{e.rd2}.
\end{eqnarray}
After a little algebra, one can show that
\begin{equation}
f_1(x_1,x_2; e, \mu) + f_2(x_2,x_1; e, \mu) = \frac{1}{2} \mu (b-c) (1-3x_1),
\label{e.dual}
\end{equation}
which becomes zero as $\mu$ vanishes. Note that $x_1$ and $x_2$
exchange their positions when they are arguments of $f_2$ in Eq.~\eqref{e.dual}.
If we set $\mu=0$ and define $\tau \equiv -t$, therefore,
\begin{eqnarray}
\frac{dx_1}{d\tau} &=& -\frac{dx_1}{dt} = -f_1 (x_1,x_2;e,0) = f_2(x_2,x_1;e,0)\\
\frac{dx_2}{d\tau} &=& -\frac{dx_2}{dt} = -f_2 (x_1,x_2;e,0) = f_1(x_2,x_1;e,0)
\end{eqnarray}
By introducing $X_1 \equiv x_2$ and $X_2 \equiv x_1$, we find that
\begin{eqnarray}
\frac{dX_1}{d\tau} &=& f_1(X_1,X_2;e,0)\\
\frac{dX_2}{d\tau} &=& f_2(X_1,X_2;e,0),
\end{eqnarray}
which recovers the original dynamics. In other words, the dynamics is dual under
time reversal and exchange of $x_1$ and $x_2$. Suppose that we have observed
a trajectory $(x_1(t), x_2(t))$ under RD with $\mu=0$. Even if we exchange the
names of AllC and AllD populations and trace the trajectory backward in time, we
will obtain a valid trajectory governed by the same RD due to the duality
(Fig.~\ref{f.dual}). As a consequence, for a given fixed point (FP) $(x_1,x_2)$,
there must be a mirror FP $(x_2,x_1)$. Furthermore, the duality also imposes a
constraint on the stability: If one is stable, for example, the other must be
unstable. Suppose that RD has a single FP. We then have to conclude that $x_1 =
x_2$ because $(x_1, x_2) = (x_2, x_1)$. In addition, due to the stability
constraint, it must be either a saddle or a neutrally stable point.

\begin{figure}
\includegraphics[width=0.45\textwidth]{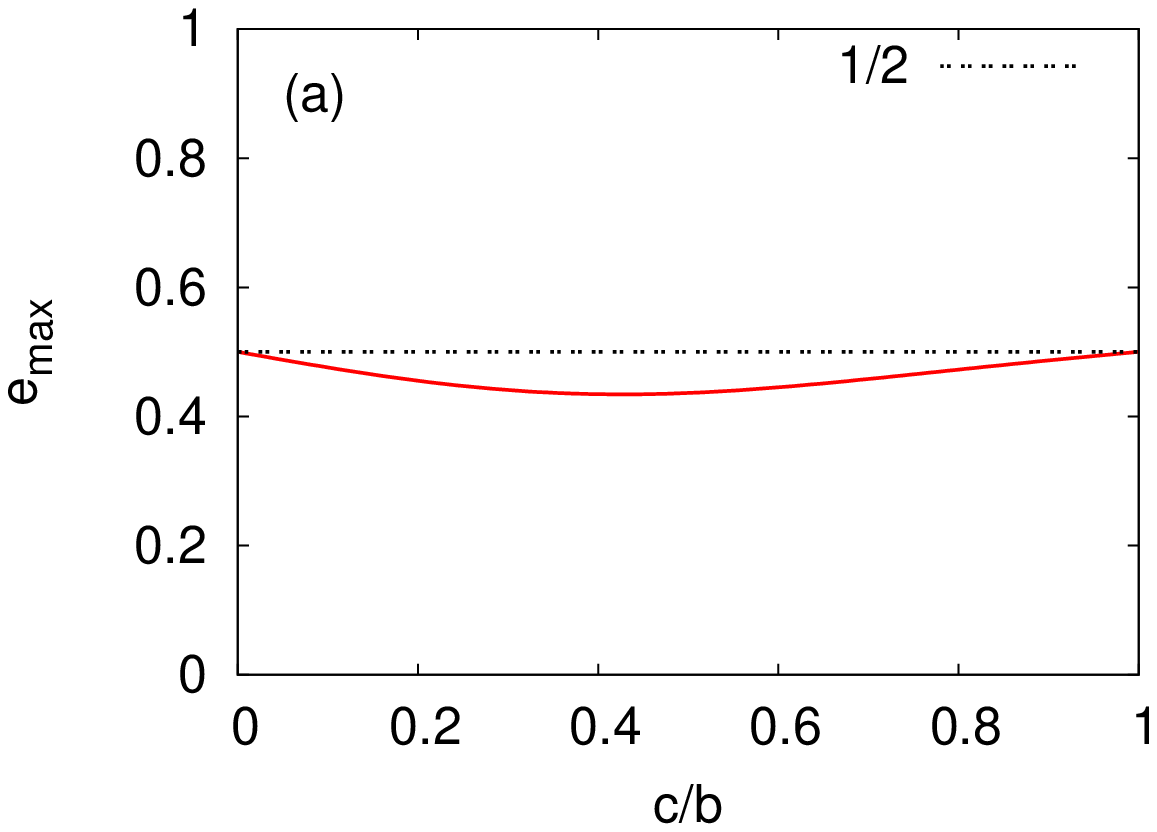}
\includegraphics[width=0.45\textwidth]{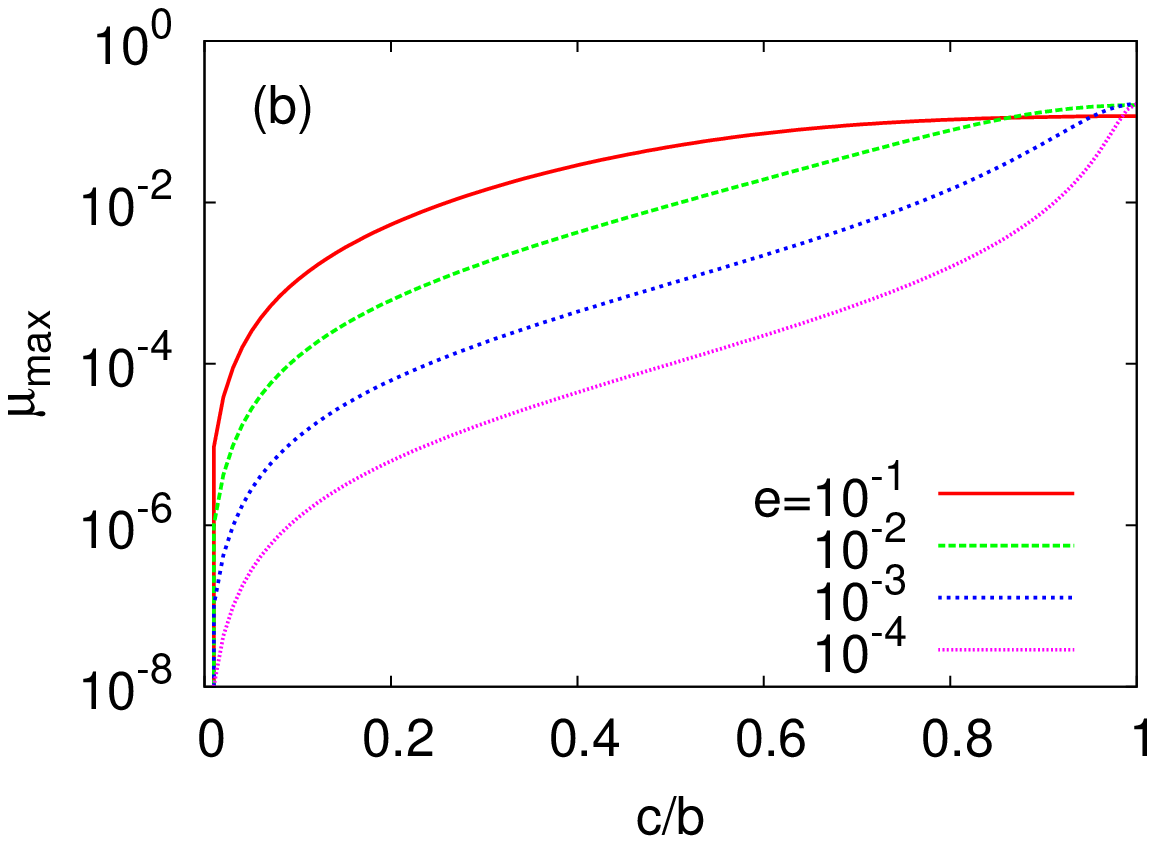}
\caption{Upper bounds of $e$ and $\mu$ for both FP$_2^\ast$ and
FP$_4^\ast$ in Eq.~\eqref{e.fps} to be real, under the assumption that $\mu \ll
e \ll 1$. (a) The inequality for $e$ [Eq.~\eqref{e.emax}] is always satisfied
for $e \ll 1$. (b) If $\mu > \mu_{\max}$, FP$_2^\ast$ and FP$_4^\ast$ disappear
from the real domain.}
\label{f.max}
\end{figure}

The question is the number of FP's in this dynamics.
When $\mu = 0$, it is relatively easy to calculate each FP:
\begin{equation}
(x_1,x_2) =
\left\{
\begin{array}{lcl}
(1,0) &\equiv& \mbox{FP}_1\\
(0,1) &\equiv& \mbox{FP}_2\\
\left( \frac{b(1-2e)-c}{(b-c)(1-2e)} , 0\right) &\equiv& \mbox{FP}_3\\
\left( 0, \frac{b(1-2e)-c}{(b-c)(1-2e)} \right) &\equiv& \mbox{FP}_4\\
(0,0) &\equiv& \mbox{FP}_5\\
\left( \frac{b(1-2e)-c}{2b(1-2e)}, \frac{b(1-2e)-c}{2b(1-2e)}
\right) &\equiv& \mbox{FP}_6
\end{array}
\right.
\label{e.fp0}
\end{equation}
If $b(1-2e) \ge c$, all these FP's are feasible, that is, all $x_i$'s
($i=1,2,3$) belong to the unit interval $[0,1]$. Otherwise, only FP$_1$, FP$_2$,
and FP$_5$ will remain available.
We assume that $e$ is small in the sense that $c < b(1-2e)$
for values of $b$ and $c$ considered in this work.
The eigenvalues and eigenvectors of the differential equation of
Eqs.~\eqref{e.rd1} and \eqref{e.rd2} with $\mu=0$ are given in
Table~\ref{t.eigen}.
\begin{table}
\caption{Eigenvalues and eigenvectors of each FP when $\mu=0$
[Eq.~\eqref{e.fp0}]. We have defined
$A \equiv c^2 - b^2(1-2e)$, $B \equiv 2bc\sqrt{1-2e}$, and $C \equiv c^2 + b^2
(1-2e)$.}
\begin{tabular*}{\hsize}{l@{\extracolsep{\fill}}cccc@{\hspace{0cm}}}\hline\hline
FP & eigenvalue & eigenvector & eigenvalue & eigenvector \\\hline
% FP1
FP$_1$ & $c(1-2e)$ & $(-1,1)$ & $ce(1-2e)$ & $(1,0)$\\
% FP2
FP$_2$ & $-c(1-2e)$ & $(-1,1)$ & $-ce(1-2e)$ & $(0,1)$ \\
% FP3
FP$_3$ &
$\frac{c(1-2e)(b-c-2be)}{b-c}$ & $(-1,1)$ &
$-\frac{ce(b-c-2be)}{b-c}$ & $(1,0)$ \\
% FP4
FP$_4$ &
$-\frac{c(1-2e)(b-c-2be)}{b-c}$ & $(-1,1)$ &
$\frac{ce(b-c-2be)}{b-c}$ & $(0,1)$ \\
% FP5
FP$_5$ & $-\frac{1}{2}(1-2e)(b-c-2be)$ & $(0,1)$
& $\frac{1}{2}(1-2e)(b-c-2be)$ & $(1,0)$ \\
% FP6
FP$_6$
& $i\frac{c\sqrt{1-2e} (b-c-2be)}{2b}$ & $(A+ Bi, C)$
& -$i\frac{c\sqrt{1-2e} (b-c-2be)}{2b}$ & $(A- Bi, C)$ \\\hline\hline
\end{tabular*}
\label{t.eigen}
\end{table}
For $\mu > 0$, we cannot find all FP's in closed forms because they are involved
with a sixth-order polynomial equation. It is more instructive to calculate them
in a perturbative way for small $\mu$.
We obtain the perturbative solution by using the Newton method, in which the
FP's for $\mu=0$ serve as trial solutions. Let us denote any of the
trial solutions as $(x_1, x_2)$, whereas the corresponding solution for
$\mu>0$ as $(x_1^\ast, x_2^\ast)$.
From the Taylor expansion around the FP:
\begin{eqnarray}
0 &=& f_1 (x_1^\ast, x_2^\ast) = f_1 (x_1,x_2) + (x_1^\ast-x_1) \frac{\partial
f_1}{\partial x_1} + (x_2^\ast - x_2) \frac{\partial f_1}{\partial x_2} +
\ldots\\
0 &=& f_2 (x_1^\ast, x_2^\ast) = f_2 (x_1,x_2) + (x_1^\ast-x_1) \frac{\partial
f_2}{\partial x_1} + (x_2^\ast - x_2) \frac{\partial f_2}{\partial x_2} +
\ldots,
\end{eqnarray}
we observe that
\begin{equation}
\begin{pmatrix}
x_1^\ast\\
x_2^\ast
\end{pmatrix}
\approx
\begin{pmatrix}
x_1\\
x_2
\end{pmatrix}
-
\begin{pmatrix}
\frac{\partial f_1}{\partial x_1} & \frac{\partial f_1}{\partial x_2}\\
\frac{\partial f_2}{\partial x_1} & \frac{\partial f_2}{\partial x_2}
\end{pmatrix}^{-1}
\begin{pmatrix}
f_1 (x_1, x_2)\\
f_2 (x_1, x_2)
\end{pmatrix}.
\end{equation}
The resulting expressions for $\mu > 0$ are the followings:
\begin{equation}
(x_1^\ast,x_2^\ast) \approx
\left\{
\begin{array}{lcl}
(1,0) &+& \mu \left( \frac{(b-c)(1-e^2)}{2ce(1-2e)}, \frac{-(b-c)(1-e)}{2c(1-2e)} \right)\\
(0,1) &+& \mu \left( \frac{(b-c)e}{2c(1-2e)}, \frac{-(b-c)(1+e)}{2c(1-2e)}
\right)\\
\left( \frac{b(1-2e)-c}{(b-c)(1-2e)} , 0\right) &+& \mu \left(
\frac{[b-c-3(b+c)e] [2bc(1-e+e^2)-(b^2+c^2)(1-e)]}{2c(b-c)e(1-2e)(b-c-2be)},
-\frac{(b-c)^2(1-e)-2bce^2}{2c(1-2e)(b-c-2be)} \right)\\
\left( 0, \frac{b(1-2e)-c}{(b-c)(1-2e)} \right) &+& \mu \left(
\frac{e[(b-c)^2+2bce]}{2c(1-2e)(b-c-2be)},
\frac{[(b-c)^2+2bce][b-c-3(b+c)e]}{2c(b-c)(1-2e)(b-c-2be)} \right)\\
(0,0) &+& \mu \left( -\frac{(b-c)}{2(1-2e)(b-c-2be)},
\frac{(b-c)}{2(1-2e)(b-c-2be)} \right)\\
\left( \frac{b(1-2e)-c}{2b(1-2e)}, \frac{b(1-2e)-c}{2b(1-2e)} \right) &+& \mu
\left( \frac{b(b-c)(b-3c-2be)}{4c^2(1-2e)(b-c-2be)},
-\frac{b(b-c)(b-3c-2be)}{4c^2(1-2e)(b-c-2be)} \right).
\end{array}
\right.
\label{e.fps}
\end{equation}
Recall that we are concerned with a parameter region of $c < b(1-2e)$.
We discard the first, third, and fifth solutions because they admit negative
fractions in this region. We will denote the other three as FP$_2^\ast$,
FP$_4^\ast$, and FP$_6^\ast$, respectively. Some of them can also be unfeasible,
however,
because an implicit assumption behind Eq.~\eqref{e.fps} is that the perturbed
solutions still exist in the real domain, which may not be always true.
It turns out that FP$_2^\ast$ and FP$_4^\ast$ can be complex unless we
restrict the ranges of $e$ and $\mu$. In Appendix, we derive the following set
of inequalities to make FP$_2^\ast$ and FP$_4^\ast$ real, provided
that $\mu \ll e \ll 1$:
\begin{eqnarray}
e &\lesssim& e_{\max} \equiv \frac{-b^2+3bc+4c^2 +
\sqrt{9b^4-18b^3c+37b^2c^2-44bc^3+52c^4}}{2(2b^2 + bc + 9c^2)}\label{e.emax}\\
\mu &\lesssim& \mu_{\max} \equiv \frac{c^2 e (1-2e)}{(b-c)^2 -e
(b^2-3bc-4c^2) - e^2 (2b^2 -bc - 9c^2)}.\label{e.mumax}
\end{eqnarray}
We plot the upper bounds $e_{\max}$ and $\mu_{\max}$ in Fig.~\ref{f.max}.
From Fig.~\ref{f.max}(a), we see that the first inequality
is always satisfied as long as $e \ll 1$. For given $e$ and $\mu$,
one can solve $\mu > \mu_{\max}$ to estimate the range of $c$ that makes
FP$_2^\ast$ and FP$_4^\ast$ complex, leaving only FP$_6^\ast$ as a possible
outcome.
The point is that FP$_6^\ast$, the last one in Eq.~\eqref{e.fps}, is the most
robust one which remains feasible over the range of $c$ under consideration.
To tell if it is actually accessible, we should analyze its stability.
Table~\ref{t.eigen} shows that it is neutrally stable at $\mu = 0$ because
its eigenvalues are purely imaginary. Let us denote the eigenvalues as
$\lambda_6^{\pm}$, where $\pm$ means the sign in front.
If we introduce small yet positive $\mu$, they begin to contain a real part with
a magnitude of $O(\mu)$:
\begin{equation}
\mbox{Re}(\lambda_6^{\pm}) \approx -\mu\frac{(b-c)[(b-c)^2+2c^2]}{4c(b-c-2be)},
\label{e.stable}
\end{equation}
which is negative in our parameter region. It means that FP$_6^\ast$ will be
stable in the presence of mutation so that nearby trajectories will be attracted
to that point. If $c < b(1-2e)/3$, the correction is positive for $x_1$ and
negative for $x_2$. Mutation breaks the duality between cooperators and
defectors, and it does in a way that favors and stabilizes cooperation.

\section{Numerical Results}

\begin{figure}
\includegraphics[width=0.45\textwidth]{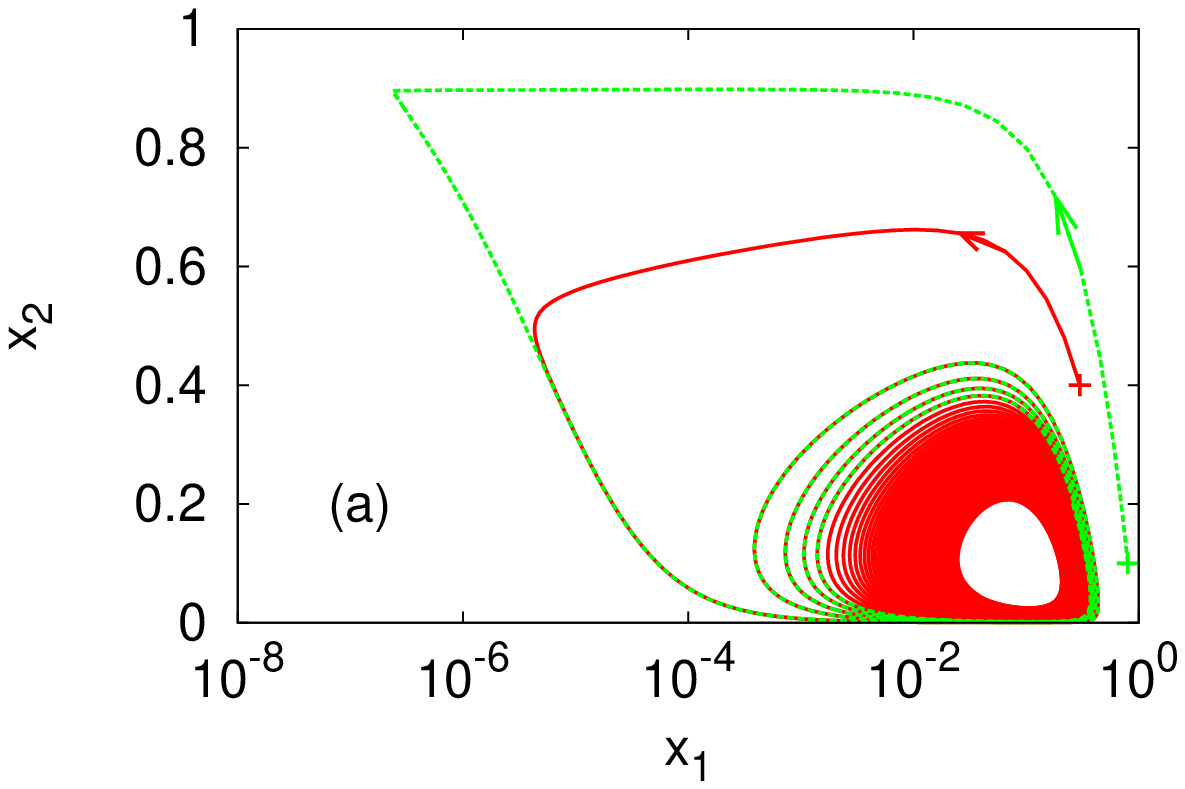}
\includegraphics[width=0.45\textwidth]{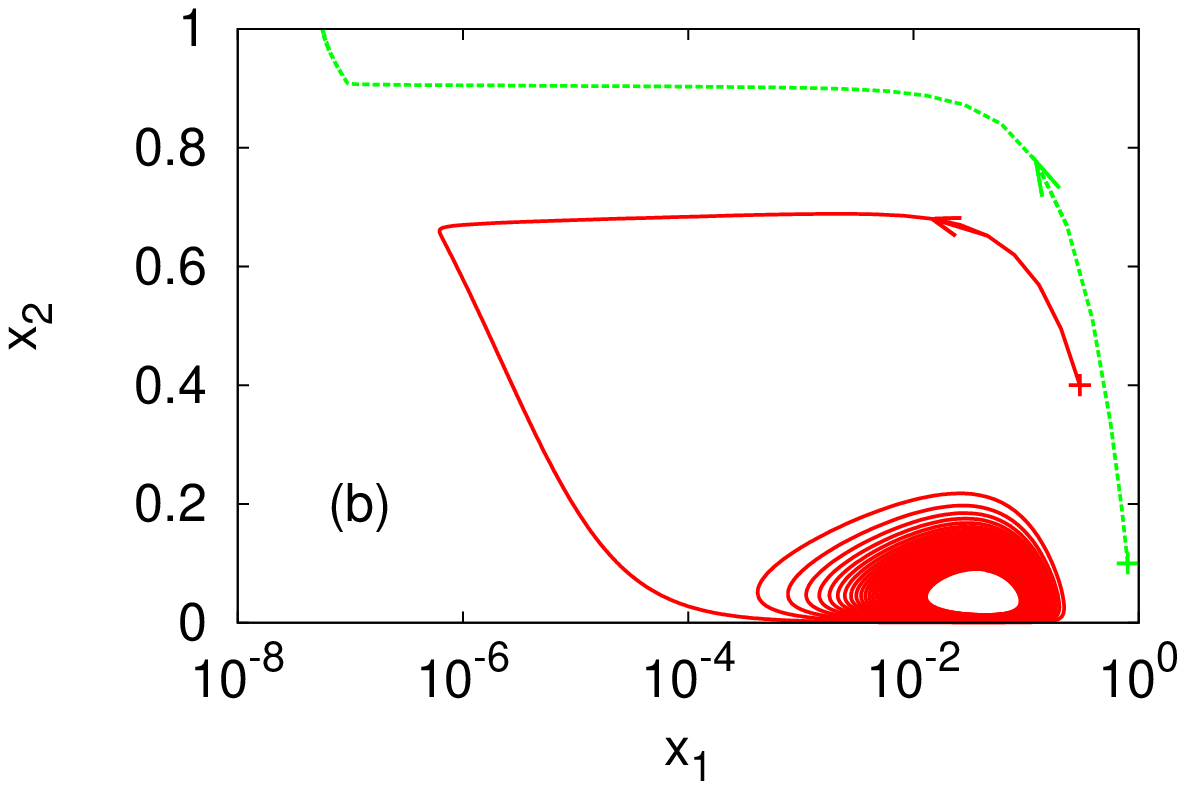}
\caption{Numerical integration of RD for (a) $c=0.8$ and (b) $c=0.9$, with
$e=10^{-2}$ and $\mu=10^{-4}$.
In each panel, we plot trajectories for two different initial conditions,
$(x_1,x_2) = (0,3,0.4)$ and $(0.8,0.1)$, represented by the crosses. When
$c=0.8$, both converge to FP$_6^\ast$ with $x_1 = x_2 \approx (1-c/b)/2$.
On the other hand, if $c=0.9$, one of them is attracted to FP$_2^\ast \approx
(0,1)$.}
\label{f.traj}
\end{figure}

\begin{figure}
\includegraphics[width=0.45\textwidth]{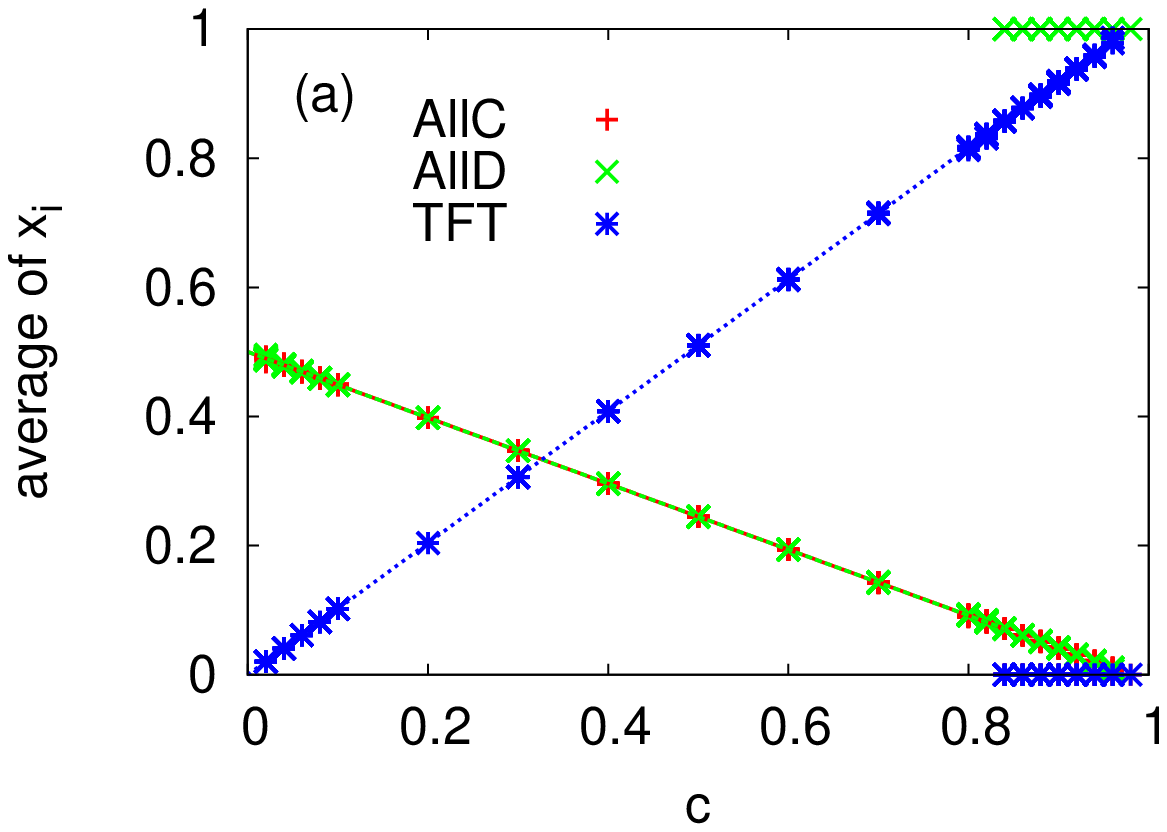}
\includegraphics[width=0.45\textwidth]{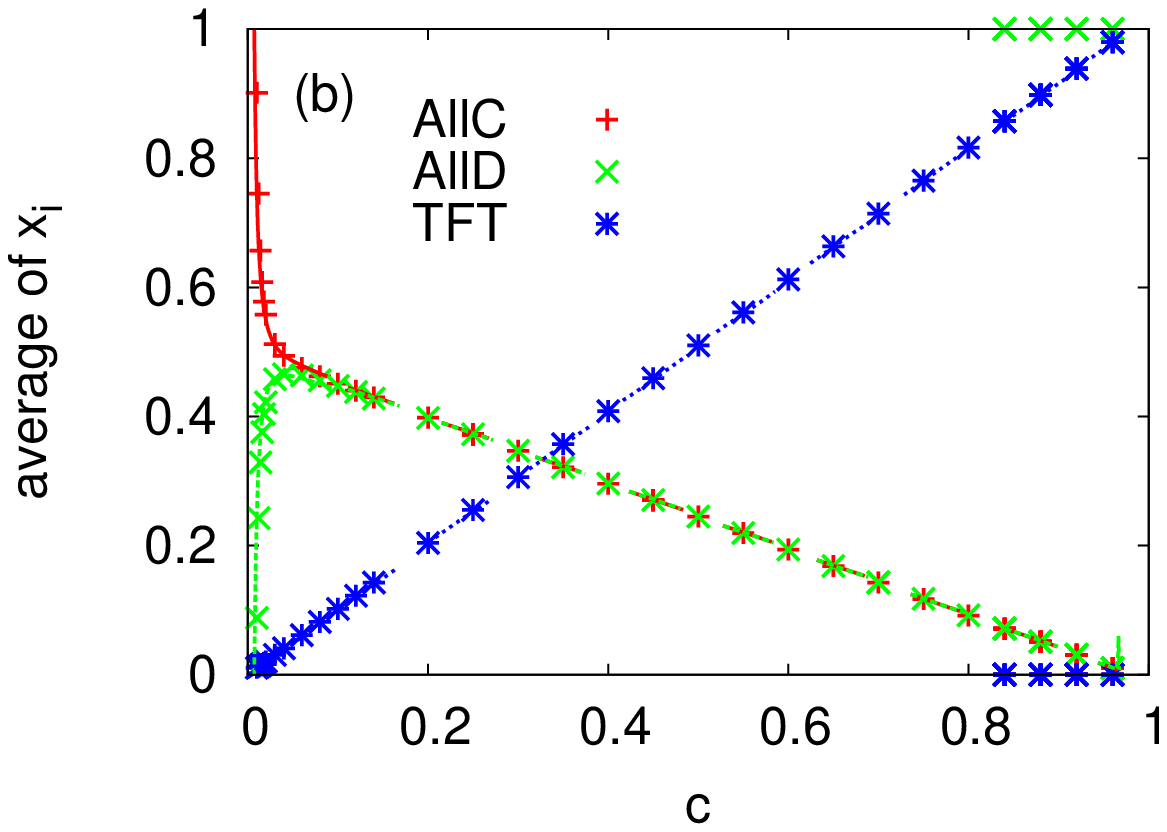}
\caption{$\av{x}_i$ as a function of $c$ from numerical integration of RD.
Fixing $b=1$ and $e=10^{-2}$, we try two different values for the mutation
rate (a) $\mu = 0$ and (b) $\mu = 10^{-4}$, respectively. The lines represent
the last solution in Eq.~\eqref{e.fps}, denoted as FP$_6^\ast$.
We have checked an exhaustive list of initial fractions with mesh size
$\frac{1}{10}$ (see text). For some $c$, $\av{x}_i$ looks multi-valued because
the system approaches different attractors depending on the initial condition.
}
\label{f.rd3e}
\end{figure}

We have performed numerical calculations to check our analytic
calculations in the previous section.
We fix $b$ as unity without loss of generality.
We have chosen $e = 10^{-2}$, so the inequality $b(1-2e)>c$ is satisfied for
$0<c<0.98$.
Integrating Eq.~(\ref{e.rd}) from an initial condition, we remove transient
behavior and calculate the time averages of $x_\alpha$ defined as follows:
\beqa
 \av{x}_\alpha \a= \lim_{T\to\infty} \frac{1}{T-T_0} \int_{T_o}^T\,
x_\alpha\,dt,
\label{e.avxi}
\eeqa
where $T_0$ is transient time. Note that the dynamics may have multiple
attractors:
Figures~\ref{f.traj}(a) and (b) show numerical integration of RD when $c=0.8$
and $c=0.9$, respectively. Sometimes every initial condition leads to the
same result on average [Fig.~\ref{f.traj}(a)]. Then, we can express any of
$\av{x}_i$'s ($i=1,2,3$) as a function of $c$. However, if this is not the
case, as illustrated in Fig.~\ref{f.traj}(b), we have to test many different
initial conditions, and the resulting $\av{x}_i$ will be multi-valued for given
$c$.
To sample the initial condition, we use an exhaustive search with mesh size
$\frac{1}{10}$. That is, we check initial conditions of (AllC, AllD, TFT) =
$(\frac{1}{10}, \frac{1}{10}, \frac{8}{10}), (\frac{1}{10}, \frac{2}{10},
\frac{7}{10}), \ldots, (\frac{8}{10}, \frac{1}{10}, \frac{1}{10})$.

In Fig.~\ref{f.rd3e}(a), we have depicted how $\av{x}_i$ depends on $c$
when $\mu=0$. For $c \lesssim c_b \approx 0.8$, every initial condition yields
the same result in the long run, which agrees with FP$_6^\ast$ very well. For
$c>c_b$, the system is bistable and we get two different pairs of
$(\av{x}_1,\av{x}_2)$. One of them still agrees with FP$_6$, while the other
coincides with FP$_2 = (0,1)$.
Figure~\ref{f.rd3e}(b) shows the case of $\mu=10^{-4}$, for which the overall
behavior is essentially same as in Fig.~\ref{f.rd3e}(a) except at small $c$.
This is because $\Delta x_i = x_i^* - x_i$ is of
$O(\mu/c^2)$ in FP$_6^\ast$, as presented in Eq.~\eqref{e.fps}. Hence the
correction due to $\mu = 10^{-4}$ is visible only for $c \sim O(\mu^{1/2}) =
O(10^{-2})$. Interestingly, the correction term in
FP$_6^\ast$ has singularity at $c=0$, whereas the fractions $x_1^\ast$ and
$x_2^\ast$ must be bounded. For this reason, our perturbative analysis obviously
breaks down as $c \rightarrow 0$. Having said that, the agreement in
Fig.~\ref{f.rd3e} is truly remarkable.
On the other hand, the existence of multiple FP's is detected
only at $c > c_b$, although Eq.~\eqref{e.mumax} is satisfied for $c
\gtrsim 0.09$ according to our parameters $b=1$, $e=10^{-2}$ and $\mu=10^{-4}$.
It suggests that FP$_5^\ast$ has small basins of attraction, compared to our
mesh size: The population is mostly occupied by AllD at FP$_4^\ast$, but it
cannot be sustained unless the TFT population is very small.

\section{Discussion and Summary}

Before concluding this work, let us consider how our observation can be
generalized. In fact, the structure of RD seems to be crucial for the existence
of such duality: We have also checked the same strategy set with the Moran
process for a finite
population~\cite{baek:scirep:2016,nowak:nature:2004,taylor:bmb:2004,jeong:jtb:2014},
but we do not find such a symmetry between AllC and AllD (not shown).
In this sense, the duality between AllC and AllD is not universal.
Another related question is whether other sets of strategies can also
exhibit the same kind of duality, provided that RD governs time evolution.
To be more specific, let $i$, $j$, and $k$ be three different strategies, i.e.,
$i \neq j$, $j \neq k$, and $i \neq k$ with fractions $x_i$, $x_j$, and $x_k$,
respectively. Just as Eqs.~\eqref{e.rd1} to \eqref{e.dual}, the duality means
that $f_i (x_i, x_j) + f_j(x_j, x_i) = 0$ when mutation is absent. It turns out
that our strategy set is not the only possibility:
%We have found $40$ such sets within the scope of memory-one strategies.
One particularly interesting case of duality is such that $i=$AllC and $j=$AllD
as before, whereas $k=$TFT is replaced by anti-TFT, which is a reactive strategy
described as $(P_C, P_D) = (0,1)$. Therefore, the duality alone does not
determine which strategy set one should work with. We believe that one should
first define a larger set of strategies from a general constraint, such as
memory length, and then pick up the most important ones therein {\it a
posteriori}. Along this line, the choice of AllC, AllD, and TFT becomes
most meaningful in an environment with a moderate value of $c$, where TFT
occupies a substantial fraction of the population and other surviving strategies
can be classified into cooperative and non-cooperative ones.

To summarize, we have investigated IPD of three representative strategies, AllC,
AllD, and TFT, by analyzing RD as a dynamical system. We have shown duality
between the fractions of cooperators and defectors in the absence of mutation.
The effects of small positive $\mu$ have been studied in a perturbative manner:
Mutation enhances cooperation if $c/b \lesssim 1/3$ and stabilizes the
corresponding fixed point. The enhancement becomes significant especially for
$c/b < O(\mu^{1/2})$.
These results have been confirmed by numerical calculations.
Our finding implies that evolutionary dynamics may have a variety of emergent
symmetries. According to this picture, a defecting population can be viewed as a
cooperating population traveling backward in time, and vice versa, in the
presence of TFT.

\acknowledgments
S.K.B. was supported by Basic Science Research Program through the National
Research Foundation of Korea (NRF) funded by the Ministry of Science, ICT and
Future Planning (NRF-2017R1A1A1A05001482).

\appendix
\section*{Derivation of Eq.~\eqref{e.mumax}}
As $\mu$ increases from zero, FP$_2^\ast$ and FP$_4^\ast$, the second and fourth
solutions in Eq.~\eqref{e.fps}, become complex via a saddle-node bifurcation.
When the bifurcation point is approached, the deviation of $x_1$ from zero is
entirely due to $\mu$, whereas the deviation of $x_2$ from unity has a
contribution from $e$. It is therefore plausible to assume that $x_1 \ll 1-x_2$.
We thus expand $f_1(x_1, x_2;\mu)$ and $f_2(x_1, x_2;\mu)$ in Eqs.~\eqref{e.rd1}
and \eqref{e.rd2} around $(x_1,x_2) = (0,1)$ to the linear order in $x_1$ and to
the second order in $(1-x_2)$.

By solving $d{x}_1/dt = d{x}_2/dt = 0$ in this set of reduced equations,
we explicitly obtain approximate formulas for the FP's. They contain a common
factor, which we denote as $\sqrt{g(b,c,e,\mu)}$, and this is the only factor
that can make the FP's complex. We simplify $g$ by expanding it to the linear
order in $\mu$, and calculate the conditions for it to be non-negative. One of
the resulting sets of conditions is written in Eqs.~\eqref{e.emax} and
\eqref{e.mumax}. The other has been discarded because it is valid only for a
high error rate.

%\bibliographystyle{apsrev4-1}
%\bibliography{dual}

\begin{thebibliography}{12}%
\makeatletter
\providecommand \@ifxundefined [1]{%
 \@ifx{#1\undefined}
}%
\providecommand \@ifnum [1]{%
 \ifnum #1\expandafter \@firstoftwo
 \else \expandafter \@secondoftwo
 \fi
}%
\providecommand \@ifx [1]{%
 \ifx #1\expandafter \@firstoftwo
 \else \expandafter \@secondoftwo
 \fi
}%
\providecommand \natexlab [1]{#1}%
\providecommand \enquote  [1]{``#1''}%
\providecommand \bibnamefont  [1]{#1}%
\providecommand \bibfnamefont [1]{#1}%
\providecommand \citenamefont [1]{#1}%
\providecommand \href@noop [0]{\@secondoftwo}%
\providecommand \href [0]{\begingroup \@sanitize@url \@href}%
\providecommand \@href[1]{\@@startlink{#1}\@@href}%
\providecommand \@@href[1]{\endgroup#1\@@endlink}%
\providecommand \@sanitize@url [0]{\catcode `\\12\catcode `\$12\catcode
  `\&12\catcode `\#12\catcode `\^12\catcode `\_12\catcode `\%12\relax}%
\providecommand \@@startlink[1]{}%
\providecommand \@@endlink[0]{}%
\providecommand \url  [0]{\begingroup\@sanitize@url \@url }%
\providecommand \@url [1]{\endgroup\@href {#1}{\urlprefix }}%
\providecommand \urlprefix  [0]{URL }%
\providecommand \Eprint [0]{\href }%
\providecommand \doibase [0]{http://dx.doi.org/}%
\providecommand \selectlanguage [0]{\@gobble}%
\providecommand \bibinfo  [0]{\@secondoftwo}%
\providecommand \bibfield  [0]{\@secondoftwo}%
\providecommand \translation [1]{[#1]}%
\providecommand \BibitemOpen [0]{}%
\providecommand \bibitemStop [0]{}%
\providecommand \bibitemNoStop [0]{.\EOS\space}%
\providecommand \EOS [0]{\spacefactor3000\relax}%
\providecommand \BibitemShut  [1]{\csname bibitem#1\endcsname}%
\let\auto@bib@innerbib\@empty
%</preamble>
\bibitem [{\citenamefont {Axelrod}(1984)}]{axelrod:book:1994}%
  \BibitemOpen
  \bibfield  {author} {\bibinfo {author} {\bibfnamefont {R.}~\bibnamefont
  {Axelrod}},\ }\href@noop {} {\emph {\bibinfo {title} {The Evolution of
  Cooperation}}}\ (\bibinfo  {publisher} {Basic Books},\ \bibinfo {address}
  {New York},\ \bibinfo {year} {1984})\BibitemShut {NoStop}%
\bibitem [{\citenamefont {Imhof}\ \emph {et~al.}(2005)\citenamefont {Imhof},
  \citenamefont {Fudenberg},\ and\ \citenamefont {Nowak}}]{imhof:pnas:2005}%
  \BibitemOpen
  \bibfield  {author} {\bibinfo {author} {\bibfnamefont {L.~A.}\ \bibnamefont
  {Imhof}}, \bibinfo {author} {\bibfnamefont {D.}~\bibnamefont {Fudenberg}}, \
  and\ \bibinfo {author} {\bibfnamefont {M.~A.}\ \bibnamefont {Nowak}},\
  }\href@noop {} {\bibfield  {journal} {\bibinfo  {journal} {Proc. Natl. Acad.
  Sci. USA}\ }\textbf {\bibinfo {volume} {102}},\ \bibinfo {pages} {10797}
  (\bibinfo {year} {2005})}\BibitemShut {NoStop}%
\bibitem [{\citenamefont {Brandt}\ and\ \citenamefont
  {Sigmund}(2006)}]{brandt:jtb:2006}%
  \BibitemOpen
  \bibfield  {author} {\bibinfo {author} {\bibfnamefont {H.}~\bibnamefont
  {Brandt}}\ and\ \bibinfo {author} {\bibfnamefont {K.}~\bibnamefont
  {Sigmund}},\ }\href@noop {} {\bibfield  {journal} {\bibinfo  {journal} {J.
  Theor. Biol.}\ }\textbf {\bibinfo {volume} {239}},\ \bibinfo {pages} {183}
  (\bibinfo {year} {2006})}\BibitemShut {NoStop}%
\bibitem [{\citenamefont {Toupo}\ \emph {et~al.}(2014)\citenamefont {Toupo},
  \citenamefont {Rand},\ and\ \citenamefont {Strogatz}}]{toupo:ijbc:2014}%
  \BibitemOpen
  \bibfield  {author} {\bibinfo {author} {\bibfnamefont {D.~F.~P.}\
  \bibnamefont {Toupo}}, \bibinfo {author} {\bibfnamefont {D.~G.}\ \bibnamefont
  {Rand}}, \ and\ \bibinfo {author} {\bibfnamefont {S.~H.}\ \bibnamefont
  {Strogatz}},\ }\href@noop {} {\bibfield  {journal} {\bibinfo  {journal} {Int.
  J. Bifurcat. Chaos}\ }\textbf {\bibinfo {volume} {24}},\ \bibinfo {pages}
  {1430035} (\bibinfo {year} {2014})}\BibitemShut {NoStop}%
\bibitem [{\citenamefont {Baek}\ \emph {et~al.}(2016)\citenamefont {Baek},
  \citenamefont {Jeong}, \citenamefont {Hilbe},\ and\ \citenamefont
  {Nowak}}]{baek:scirep:2016}%
  \BibitemOpen
  \bibfield  {author} {\bibinfo {author} {\bibfnamefont {S.~K.}\ \bibnamefont
  {Baek}}, \bibinfo {author} {\bibfnamefont {H.-C.}\ \bibnamefont {Jeong}},
  \bibinfo {author} {\bibfnamefont {C.}~\bibnamefont {Hilbe}}, \ and\ \bibinfo
  {author} {\bibfnamefont {M.~A.}\ \bibnamefont {Nowak}},\ }\href@noop {}
  {\bibfield  {journal} {\bibinfo  {journal} {Sci. Rep.}\ }\textbf {\bibinfo
  {volume} {6}},\ \bibinfo {pages} {25676} (\bibinfo {year}
  {2016})}\BibitemShut {NoStop}%
\bibitem [{\citenamefont {Nowak}\ and\ \citenamefont
  {Sigmund}(1989)}]{nowak:jtb:1989}%
  \BibitemOpen
  \bibfield  {author} {\bibinfo {author} {\bibfnamefont {M.~A.}\ \bibnamefont
  {Nowak}}\ and\ \bibinfo {author} {\bibfnamefont {K.}~\bibnamefont
  {Sigmund}},\ }\href@noop {} {\bibfield  {journal} {\bibinfo  {journal} {J.
  Theor. Biol.}\ }\textbf {\bibinfo {volume} {137}},\ \bibinfo {pages} {21}
  (\bibinfo {year} {1989})}\BibitemShut {NoStop}%
\bibitem [{\citenamefont {Nowak}\ and\ \citenamefont
  {Sigmund}(1990)}]{nowak:aam:1990}%
  \BibitemOpen
  \bibfield  {author} {\bibinfo {author} {\bibfnamefont {M.~A.}\ \bibnamefont
  {Nowak}}\ and\ \bibinfo {author} {\bibfnamefont {K.}~\bibnamefont
  {Sigmund}},\ }\href@noop {} {\bibfield  {journal} {\bibinfo  {journal} {Acta
  Appl. Math.}\ }\textbf {\bibinfo {volume} {20}},\ \bibinfo {pages} {247}
  (\bibinfo {year} {1990})}\BibitemShut {NoStop}%
\bibitem [{\citenamefont {Molander}(1985)}]{molander:jcr:1985}%
  \BibitemOpen
  \bibfield  {author} {\bibinfo {author} {\bibfnamefont {P.}~\bibnamefont
  {Molander}},\ }\href@noop {} {\bibfield  {journal} {\bibinfo  {journal} {J.
  Conflict Resolut.}\ }\textbf {\bibinfo {volume} {29}},\ \bibinfo {pages}
  {611} (\bibinfo {year} {1985})}\BibitemShut {NoStop}%
\bibitem [{\citenamefont {Imhof}\ \emph {et~al.}(2007)\citenamefont {Imhof},
  \citenamefont {Fudenberg},\ and\ \citenamefont {Nowak}}]{imhof:jtb:2007}%
  \BibitemOpen
  \bibfield  {author} {\bibinfo {author} {\bibfnamefont {L.~A.}\ \bibnamefont
  {Imhof}}, \bibinfo {author} {\bibfnamefont {D.}~\bibnamefont {Fudenberg}}, \
  and\ \bibinfo {author} {\bibfnamefont {M.~A.}\ \bibnamefont {Nowak}},\
  }\href@noop {} {\bibfield  {journal} {\bibinfo  {journal} {J. Theor. Biol.}\
  }\textbf {\bibinfo {volume} {247}},\ \bibinfo {pages} {574} (\bibinfo {year}
  {2007})}\BibitemShut {NoStop}%
\bibitem [{\citenamefont {Nowak}\ \emph {et~al.}(2004)\citenamefont {Nowak},
  \citenamefont {Sasaki}, \citenamefont {Taylor},\ and\ \citenamefont
  {Fudenberg}}]{nowak:nature:2004}%
  \BibitemOpen
  \bibfield  {author} {\bibinfo {author} {\bibfnamefont {M.~A.}\ \bibnamefont
  {Nowak}}, \bibinfo {author} {\bibfnamefont {A.}~\bibnamefont {Sasaki}},
  \bibinfo {author} {\bibfnamefont {C.}~\bibnamefont {Taylor}}, \ and\ \bibinfo
  {author} {\bibfnamefont {D.}~\bibnamefont {Fudenberg}},\ }\href@noop {}
  {\bibfield  {journal} {\bibinfo  {journal} {Nature}\ }\textbf {\bibinfo
  {volume} {428}},\ \bibinfo {pages} {646} (\bibinfo {year}
  {2004})}\BibitemShut {NoStop}%
\bibitem [{\citenamefont {Taylor}\ \emph {et~al.}(2004)\citenamefont {Taylor},
  \citenamefont {Fudenberg}, \citenamefont {Sasaki},\ and\ \citenamefont
  {Nowak}}]{taylor:bmb:2004}%
  \BibitemOpen
  \bibfield  {author} {\bibinfo {author} {\bibfnamefont {C.}~\bibnamefont
  {Taylor}}, \bibinfo {author} {\bibfnamefont {D.}~\bibnamefont {Fudenberg}},
  \bibinfo {author} {\bibfnamefont {A.}~\bibnamefont {Sasaki}}, \ and\ \bibinfo
  {author} {\bibfnamefont {M.~A.}\ \bibnamefont {Nowak}},\ }\href@noop {}
  {\bibfield  {journal} {\bibinfo  {journal} {B. Math. Biol.}\ }\textbf
  {\bibinfo {volume} {66}},\ \bibinfo {pages} {1621} (\bibinfo {year}
  {2004})}\BibitemShut {NoStop}%
\bibitem [{\citenamefont {Jeong}\ \emph {et~al.}(2014)\citenamefont {Jeong},
  \citenamefont {Oh}, \citenamefont {Allen},\ and\ \citenamefont
  {Nowak}}]{jeong:jtb:2014}%
  \BibitemOpen
  \bibfield  {author} {\bibinfo {author} {\bibfnamefont {H.-C.}\ \bibnamefont
  {Jeong}}, \bibinfo {author} {\bibfnamefont {S.-Y.}\ \bibnamefont {Oh}},
  \bibinfo {author} {\bibfnamefont {B.}~\bibnamefont {Allen}}, \ and\ \bibinfo
  {author} {\bibfnamefont {M.~A.}\ \bibnamefont {Nowak}},\ }\href@noop {}
  {\bibfield  {journal} {\bibinfo  {journal} {J. Theor. Biol.}\ }\textbf
  {\bibinfo {volume} {356}},\ \bibinfo {pages} {98} (\bibinfo {year}
  {2014})}\BibitemShut {NoStop}%
\end{thebibliography}
\end{document}